# Exploiting symmetry-mismatch to control magnetism in a ferroelastic heterostructure

Er-Jia Guo,[1,2,3,*] Ryan Desautels,[1] Dongkyu Lee,[1] Manuel A. Roldan,[4] Zhaoliang Liao,[1] Timothy Charlton,[1] Haile Ambaye,[1] Jamie Molaison,[1] Reinhard Boehler,[1,5] David Keavney,[6] Andreas Herklotz,[1,7] T. Zac Ward,[1] Ho Nyung Lee,[1] and Michael R. Fitzsimmons[1,8,*]

[1]*Oak Ridge National Laboratory, Oak Ridge, TN 37831, USA*

[2]*Beijing National Laboratory for Condensed Matter Physics and Institute of Physics, Chinese Academy of Sciences, Beijing 100190, China*

[3]*Center of Materials Science and Optoelectronics Engineering, University of Chinese Academy of Sciences, Beijing 100049, China*

[4]*Eyring Materials Center, Arizona State University, AZ 85287, United States*

[5]*Geophysical Laboratory, Carnegie Institution for Sciences, Washington DC 20005, USA*

[6]*Advanced Photon Source, Argonne National Laboratory, Argonne, Illinois 60439, USA*

[7]*Institute for Physics, Martin-Luther-University Halle-Wittenberg, Halle (Saale) 06120, Germany*

[8]*Department of Physics and Astronomy, University of Tennessee, Knoxville, TN 37996, USA*

**Corresponding Author:**

[*] *Corresponding authors: fitzsimmonsm@ornl.gov and ejguo@iphy.ac.cn*

**Abstract**

In the bulk, LaCoO$_3$ (LCO) is a paramagnet, yet in tensile-strained thin films at low temperature ferromagnetism (FM) is observed, and its origin remains unresolved. Polarized neutron reflectometry (PNR) is a powerful tool to determine the depth profiles of the structure and magnetization simultaneously and, thus, the evolution of the interfacial FM with strain can be accurately revealed. Here, we quantitatively measured the distribution of atomic density and magnetization in LCO films by PNR and found that the LCO layers near the heterointerfaces exhibit a reduced magnetization but an enhanced atomic density, whereas the film's interior (i.e., its bulk) shows the opposite trend. We attribute the nonuniformity to the symmetry mismatch at the interface, which induces a structural distortion related to the ferroelasticity of LCO. This assertion is tested by systematic application of hydrostatic pressure during the PNR experiments. The magnetization can be directly controlled at a rate of -20.4% per GPa. These results provide unique insights into mechanisms driving FM in strained LCO films while offering a tantalizing observation that tunable deformation of the CoO$_6$ octahedra in combination with the ferroelastic order parameter.

**Main text**

Ferroelastic LaCoO$_3$ (LCO) has attracted increasing attention due to the spin state transition of the cobalt ions. [1-9] The delicate interplay between the crystal field splitting ($\Delta_{CF}$) and exchange interaction ($J_{ex}$) controls the electron redistribution between the $t_{2g}$ and $e_g$ orbitals, ultimately manipulating the spin state of cobalt ions. The temperature dependent magnetic susceptibility of bulk LCO shows the spin state configuration of Co$^{3+}$ transits from a low spin (LS, $t_{2g}^6$, $S = 0$) state to an intermediate- (IS, $t_{2g}^5 e_g^1$, $S = 1$) or a high-spin state (HS, $t_{2g}^4 e_g^2$, $S = 2$) with increasing temperature. [5-9] Early work reported long-range ferromagnetism (FM) to emerge at low temperatures when the LCO was grown as thin films under tensile strain. [10-16] The origin of FM in tensile-strained LCO films remains the subject of considerable debate. The debate arises from the mechanism governing the magnetism. Previously, oxygen vacancy ordering was proposed as the origin of FM order. [17, 18] In the presence of oxygen vacancies, Co$^{3+}$ will change valence into HS Co$^{2+}$ with electron configuration of $t_{2g}^5 e_g^2$, $S = 3/2$, thus producing macroscopic long-range FM due to the excess electrons in the Co $d$ orbitals. The existence of oxygen vacancies and concomitant change of valence of some Co$^{3+}$ to Co$^{2+}$ was attributed to the appearance of dark stripes in the scanning transmission electron microscopy (STEM) images. [17, 18] However, the existence of Co$^{2+}$ in as-grown LCO films has been heavily questioned because extrinsic factors, including radiation damage from milling processes to prepare TEM specimens and imaging of them were observed to affect oxygen stoichiometry.[17-20] Furthermore, X-ray absorption spectroscopy (XAS) and spectroscopic ellipsometry measurements [21, 22] provide compelling evidence that as-grown LCO thin films were stoichiometric without detectable oxygen vacancies. An alternative mechanism attributed FM to strain-induced atomic displacements to the ferroelastic order parameter of LCO. [21] Theoretical calculations supported this interpretation.[22] Tetragonally distorted CoO$_6$ octahedra have non-zero spins, while the monoclinically distorted $CoO_6$ octahedra possess zero spin. According to this prediction, the magnetization in the LCO films might be nonuniform, *i. e.* the spin state of Co ions depends on the local structural distortion. However, until now, the crucial link between the structure and magnetism supporting the second proposal as opposed to the O-vacancy model is lacking. Therefore, there is a compelling need to directly measure the coupling of the local magnetization in an as-grown LCO film versus applied stress.

Here, we report on the depth profile of magnetization of a LCO thin film obtained from polarized neutron reflectometry (PNR) as a function of hydrostatic pressure. We find the distributions of atomic density and magnetization within a single LCO layer are nonuniform and exhibits a linear decrease as hydrostatic pressure increases, suggesting the contraction of $CoO_6$ octahedra facilitates depopulation of the HS $Co^{3+}$.

High-quality LCO film with a thickness of 40 unit cells (u.c.) capped with a STO layer (40 u.c.) was grown on a $TiO_2$-terminated $SrTiO_3$ (STO) (001) substrate by pulsed laser epitaxy (Supplementary Note and Fig. S1). Magnetic measurements show the LCO film has a Curie temperature ($T_C$) ~ 80 K, (Fig. S1) consistent with previous observations.[17-21] A magnetic hysteresis loop was observed, indicating long-range FM ordering of our LCO film at low temperatures. XAS measurements were performed on the as-grown LCO sample in the bulk-sensitive fluorescence yield (FY) mode (Fig. S2). From the XAS data, we did not observe the fingerprint of $Co^{2+}$. By comparing the spectral line shape of reference data, we confirmed the stoichiometry of our LCO film to be $LaCo^{3+}O_3$. The XAS result places an upper limit of 2% for the oxygen vacancy concentration—much less than was reported previously (~ 30%).[17, 18, 21] The oxygen vacancy concentration and concomitant change of Co valence previously observed in TEM samples is not occurring in our as-grown capped LCO film. Yet, our as-grown film exhibits FM (described below).

The chemical depth profile was obtained from fitting a model to the X-ray reflectivity (XRR) data using GenX, [23] as shown in Fig. 1a. Different models were tried to get the best fits to the XRR data (Fig. S3). The best fitting model is one with atomic densities of the two LCO interface layers in contact to STO being the same, yet different from the atomic density of the film bulk. The chemical depth profile was constrained for the analysis of the neutron reflectivity. From the X-ray scattering length density (SLD) (Fig. 1b), we find that the electron density within the LCO layer was nonuniform (Table I). The SLDs of the interfacial LCO layers in proximity to the STO have an SLD that is within 0.1% of being the same as that of bulk LCO. [8,10,24] The SLD of the LCO film's interior is smaller by 1.8 % compared to those of the interfacial layers (and bulk LCO). The reduced SLD for the film interior may be a consequence of ferroelastic domain walls, which like grain and twin boundaries, are likely regions of lower mass density. [25] The trend for the density of LCO at the film interior to be less than that at the interface was confirmed by XRR analysis of a STO/LCO superlattice with ultrathin LCO layers sandwiched between two STO layers (Fig. S4). Importantly, the reduced x-ray SLD of the film's interior cannot be attributed to oxygen vacancies, because even if oxygen vacancies were present at the upper limit of 2%, the x-ray SLD would be changed by at most 0.07%.

We performed scanning transmission electron microscopy (STEM) measurements on the same LCO sample after our non-destructive studies were completed. As shown in Fig. 1c and Fig.S5, layers without dark stripes, and commensurate with the dense region per XRR analysis, appear at or near both LCO/STO interfaces. The LCO interfacial layers have a thickness of 4-5

u.c. (i.e., the dark stripes terminate ~2 nm from the LCO/STO interface), which is consistent with XRR fitting results and previous work. [17-22]

We attribute inhomogeneous electron density (which mimics the atomic density profile) within the as-grown LCO layer to lattice distortion, i.e., the atomic stacking arrangements, induced by symmetry mismatch between STO and LCO and the strain field extending from the interface into the film. Bulk STO has a cubic lattice structure ($Pm\bar{3}m$ [221]), whereas the bulk LCO has a rhombohedral lattice structure ($R\bar{3}c$ [167]). [26] Thus, the LCO layer must compensate for a change of symmetry through a distortion at or near the interface. [21] Reciprocal space mapping (RSM) of the LCO films with different thicknesses confirm that an ultrathin LCO film having a thickness equal to the sum of its interface layers is distorted with a pseudotetragonal structure, whereas the 40 u.c.-thick LCO films show a lower symmetric monoclinically distorted structure (Fig. S6).

We performed PNR measurements on the LCO film to quantitatively determine the magnetization depth profile across the film thickness. The experiment setup and sample geometry are shown in Fig. 2a. The specular neutron reflectivity is plotted as a function of the wave vector transfer $q$ ($=4\pi sin\theta_i/\lambda$) and normalized to the asymptotic value of the Fresnel reflectivity $R_F = (16\pi^2/q^4)$ for the spin-up ($R^+$) and spin-down ($R^-$) polarized neutrons, where $\theta_i$ is the incident angle and $\lambda$ is the neutron wavelength. Solid lines in Fig. 2b are the best fit to our PNR data yielding a $\chi^2$ metric of 1.23. The spin asymmetry SA [$= (R^+–R^-)/(R^++R^-)$] and its corresponding fit were calculated from the PNR data and the results from modeling, as shown in Fig. 2c. The nuclear and magnetic SLD depth profiles of the LCO heterostructure are plotted in Fig. 2d and 2e, respectively. The magnetic depth profile indicates a large magnetization (0.96 ± 0.03 $\mu_B$/Co) in the LCO film bulk; however, the LCO interface layers exhibit rather small magnetizations (0.40 ± 0.05 $\mu_B$/Co). The thickness of the magnetic interfacial layer was constrained to be the same as that of the dense interface region identified by XRR (4-5 u.c.). The integral of $M(z)$ yields the thickness-averaged magnetization of the LCO film. The integral is consistent with the magnetization measured by magnetometry (Table I and Fig. S1). The observed nonuniform magnetization distribution in the LCO film is consistent with recent theoretical calculations, from which a significant reduction of the magnetic moment at the surface and interface of an LCO film was predicted. [27]

We further investigated the influence of stress on the magnetization profile of the LCO film by performing PNR experiments on the same LCO film as a function of hydrostatically applied pressure. [28-35] This method of measuring magnetization, unprecedented for PNR, on the identical sample can isolate the influence of elastic stress from other factors, *e.g.* oxygen stoichiometry, chemical inhomogeneity, and microstructure, from comparisons of physical properties obtained from different samples. We obtained the depth profile of the pressure dependent magnetization, which is not achievable by other means. Details of the technique and pressure calibration can be found in the Methods and Supplementary (Fig. S7 and Fig. S8). For our PNR data fits, we constrained the layer thickness and interface roughness of the film to be the same as fitting the ambient-pressure PNR data. The Poisson value for our film is $\nu = 0.38$.[36] We expect a ~1.4% change in film *thickness*, however, this value is within the uncertainty in film thickness we obtained from XRR. We allowed the nuclear and magnetic SLDs to vary with applied pressure. Figs. 3a-3c show the calculated SAs and their fits for hydrostatic pressures of 0, 0.45, and 1.63 GPa, respectively. The corresponding nuclear and magnetization depth profiles are shown in Fig. 4a. The results show that with increasing pressure, the nuclear SLD became larger while the magnetizations of the LCO film interior and interface regions became smaller. This reinforces the conclusion that the change of magnetization with pressure is an intrinsic property of the strained LCO film.

Previous work on bulk LCO demonstrated that the lattice volume of LCO shrinks by ~ 3% when the pressure is raised up to 2 GPa. [31-33] This value matches well with the relative increase in the nuclear SLD of our LCO film under pressure. The difference in the nuclear SLD magnitude between the LCO film bulk and the interface layers reduces from 3.2 to 1.2% with increasing pressure from 0 to 1.63 GPa. The contraction of the unit cell volume for the LCO film bulk is larger than that of the interface layer under the same magnitude of pressure. This can be understood in terms of Young's modulus of bulk STO ($Y_{STO}$ ~ 273 GPa), [36] which is much larger than that of bulk LCO ($Y_{LCO}$ ~ 150 GPa). [38-40] Therefore, a larger lattice deformation is expected for the LCO film bulk, while the response of the LCO interface layers to pressure is partially constrained by clamping with the STO substrate/capping layer.

The observed increase in the nuclear SLD upon application of hydrostatic pressure is accompanied by a dramatic decrease in the magnetization of the LCO film bulk and interface layers. Fig. 4b shows the pressure dependent magnetization for both LCO film bulk and interface layers. We find that the magnetization reduces with the ratio of – (0.17 ± 0.02) $\mu_B$/Co/GPa for the film bulk region and – (0.13 ± 0.01) $\mu_B$/Co/GPa for the interface layers. The total magnetization of the LCO film is reduced by ~ 21 % per GPa. Considering Young's modulus of LCO, we can estimate a reduction of magnetization by ~ 3.1 % per 0.1 % of strain produced by hydrostatic pressure. In addition, we repeated PNR measurements on the LCO heterostructure under the same conditions after removing pressure. Fig. 3a shows the SAs of the LCO film are nearly identical before (black dots) and after (red squares) the high-pressure experiments, indicating full recovery (non-destructive) of the magnetization in the LCO film after pressure is removed, *i.e.* the change of $M(z)$ is reversible with applied pressure and further demonstrate the reproducibility of our experimental protocol. Moreover, XRD results confirmed that the crystalline quality of the LCO film was not significantly affected after application of pressure up to 1.65 GPa (Fig. S9, Supporting Information). Previous work determined the critical value of the elastic deformation (yield strength) in bulk LCO is ~ 4 GPa. [30] Therefore, for the pressures applied in the present work, we expect elastic rather than plastic deformation of the LCO film.

Our results demonstrate that the local magnetization in the LCO film is strongly coupled to strain and the distortion it causes. Distortion of CoO$_6$ octahedral affects the Co-O bond length ($d_{Co-O}$) and Co-O-Co bond angle ($\beta_{Co-O-Co}$), which influence the balance between the $\Delta_{CF}$ and $J_{ex}$, and controls the spin state transition. [29-35] For bulk LCO, the room-temperature $d_{Co-O}$ is ~1.93 Å and $\beta_{Co-O-Co}$ is ~163.5° as determined by neutron diffraction [5] and density functional theory. [41] The in-plane lattice parameter of our fully-strained LCO film is $a_{LCO}$ = 3.905 Å, which is larger than $2d_{Co-O}$. Thus, the $\beta_{Co-O-Co}$ should be close or equal to 180°. In the present case, the dominant factor for the spin state transition of the Co$^{3+}$ is the bond length $d_{Co-O}$. From the X-ray experiment, we know that the atomic density of the LCO film bulk is smaller than that of the interface layers. Thus, the LCO film bulk has a larger u.c. volume compared to that of the LCO interface layer. This leads to a larger average $d_{Co-O}$ in an octahedral coordination in the LCO film bulk. A slight increase of $d_{Co-O}$ will dramatically decrease $\Delta_{CF}$ (~ $d_{Co-O}^{-5}$), which reduces the energy cost of occupation for the $e_g$ levels as favored by Hund exchange.[2] The increased population of $e_g$ electrons in the LCO film bulk can favor transition of the Co$^{3+}$ from a LS (Fig. 4c) to a HS state (Fig. 4d or Fig. 4e). Therefore, ordered HS Co$^{3+}$ promote FM, resulting in a larger magnetization of LCO film bulk compared to that of the interface layer. In addition, early work revealed that HS Co$^{3+}$ (0.75 Å) has a larger ionic radius than that of LS Co$^{3+}$ (0.685 Å). [5-7, 29] This fact further strengthens our analysis on the anomalous expansion of unit cell volume (*e.g.* reduced atomic density) in the film bulk, indicating that the small change of local structure in the LCO thin films could possibly trigger the spin state transition. This observation is consistent with a recent first-principle calculation result in which the local structural distortion in the cobaltite is induced by the LS state to HS state transition of $Co^{3+}$. [42]

Application of hydrostatic pressure on the LCO film reduces u.c. volume. Both in-plane and out-of-plane $d_{Co-O}$ bond lengths are shortened with increasing hydrostatic pressure. Considering the Young's modulus of LCO and STO, we calculated the $d_{Co-O}$ shrinks by ~ 1.1% for the LCO film bulk and ~ 0.6% for the interface layers (clamped by STO). This would lead to an increase of $\Delta_{CF}$, which favors the lower spin states of the $Co^{3+}$ ions relative to their higher spin states. The depopulation of $e_g$ electrons breaks the stabilization of ordered HS $Co^{3+}$, resulting in a significant reduction of magnetization in the LCO film under pressure. The pressure-induced compression of the u.c. volume is larger for the LCO film bulk than for the interface layer (the latter is partially constrained by STO) (Fig.4a). Therefore, a larger reduction of magnetization in the LCO film bulk than that of the interface layer (Fig. 4b) is expected. Our results for the LCO film are consistent with previous investigations on the pressure-induced spin state transition in bulk LCO, in which the spin state of $Co^{3+}$ shifts to a lower spin state with increasing pressure. [29-35]

In summary, the quantitative magnetization depth profile across the planar interface of an LCO film was obtained by PNR. The results demonstrate that a large magnetization exists in the LCO film bulk, but a small magnetization was found for the LCO interface layer. We attribute these differences to the symmetry mismatch at the LCO/STO interface, which induces a structural distortion that suppresses the higher spin state of $Co^{3+}$. The pressure dependence of the magnetization depth profile in the LCO film was measured by PNR using a custom-built hydrostatic pressure cell. We found the magnetization of LCO film decreases dramatically with increasing pressure at a rate of –20.4% per GPa. Application of hydrostatic pressure compresses the oxygen octahedra. This process drives a substantial increase of crystal field splitting energy and, consequently, leads to the depopulation of $e_g$ states, and the tendency to favor the LS state in $Co^{3+}$. Our work provides unique insight into the strong correlation between structural distortion and magnetic properties of cobaltite thin films with multiple spin states, providing an innovative opportunity to realize novel functional properties from complex oxide heterostructures.

**Acknowledgements**

We thank Y. Liu, F. Ye, W. Tian, N. Pradhan, C. Sohn, X. Gao, H. Jeen, and F. A. Reboredo for valuable discussions. We also thank J. Wenzel for designing the cryostat used for the high-pressure cell. This work was supported by the U.S. Department of Energy (DOE), Office of Science (OS), Basic Energy Sciences (BES), Materials Sciences and Engineering Division (sample deposition and characterization). During the manuscript revision, E.J.G. was supported by the Hundred Talent Program of Chinese Academy of Sciences. The research at ORNL's SNS (PNR measurements), which is a DOE, BES Scientific User Facility, was conducted via user proposals. This research used resources of the Advanced Photon Source, a U.S. DOE, OS User Facility operated for the DOE by Argonne National Laboratory under Contract No. DE-AC02-06CH1135.

**Figures and figure captions**

**Figure 1. Structural property of the STO/LCO heterostructure.** (**a**) X-ray reflectivity (XRR) of the STO/LCO heterostructure normalized to the asymmetric value of the Fresnel reflectivity. Solid curve is the best fit to the data (open circle). The inset shows the sample geometry and measuring schematic. (**b**) X-ray scattering length density (SLD) depth profile as a function of the distance from STO substrate. Black dashed line in the inset is the X-ray SLD without taking chemical roughness into account. The error bars of the LCO SLDs are ~ $0.3 \times 10^{-6}$ Å$^{-2}$. Dashed curve represents the X-ray SLD of stoichiometric LCO bulk ($5.23\times10^{-5}$ Å$^{-2}$). (**c**) Cross-sectional high-resolution high-angle annular dark-field (HAADF) scanning transmission electron microscopy (STEM) images of STO/LCO heterostructure grown on a STO substrate.

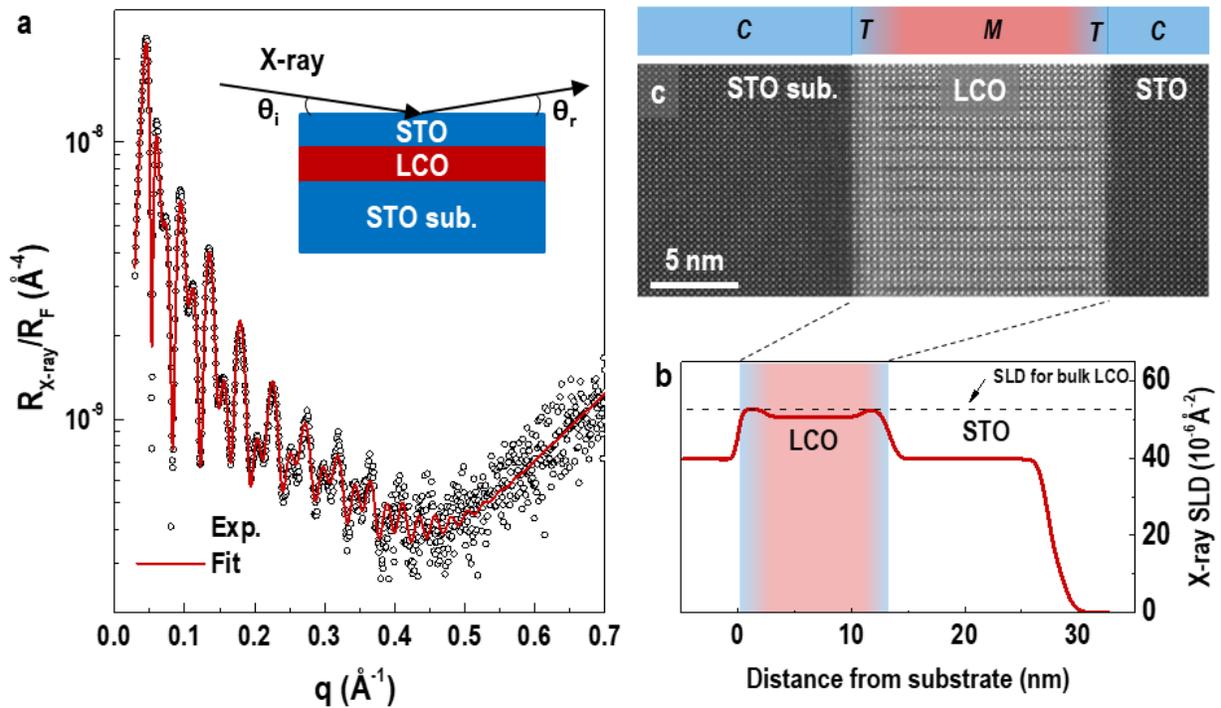

**Figure 2. PNR probing of chemical and magnetization depth profiles.** (**a**) Schematic of the PNR experimental set-up for the STO/LCO heterostructure. (**b**) Measured (symbols) and fitted (solid lines) neutron reflectivity curves for the spin-up ($R^+$) and spin-down ($R^-$) polarized neutron beams, with respect to the external magnetic field, $H$. (**c**) Fit to the spin asymmetry, SA, ratio, defined as the difference between $R^+$ and $R^-$ divided by the sum, obtained from the experimental and fitted reflectivity in (**b**). Error bars represent one standard deviation. (**d**) Nuclear and (**e**) magnetic scattering length density, SLD, depth profiles at 10 K. The error bars of the nuclear SLD within LCO is $0.031 \times 10^{-6}$ Å$^{-2}$. Dashed curve in (**d**) represents the neutron SLD of stoichiometric LCO bulk ($5.026 \times 10^{-6}$ Å$^{-2}$). The scale on the top of (**e**) shows the magnetization, $M$, of the LCO film.

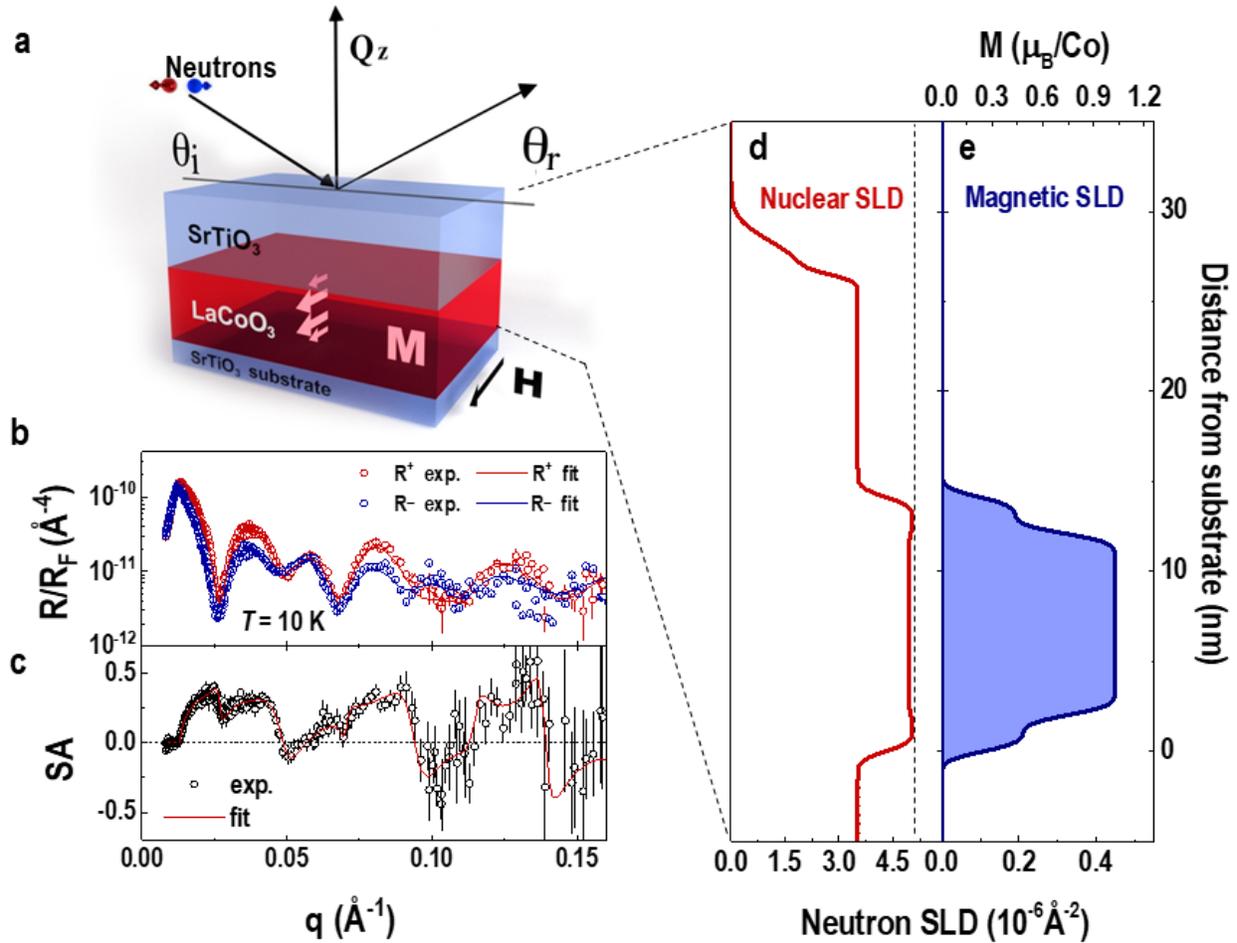

**Figure 3. Hydrostatic pressure dependent spin asymmetries and their corresponding fits.** The spin asymmetry, SA, ratios (symbols) and fitted (solid lines). The PNR measurements were performed under hydrostatic pressures of (**a**) 0, (**b**) 0.45, and (**c**) 1.63 GPa at 25 K with a magnetic field of 0.5 T. The magenta square symbols in (**a**) represent the data measured after the hydrostatic pressure is reduced to 0 GPa.

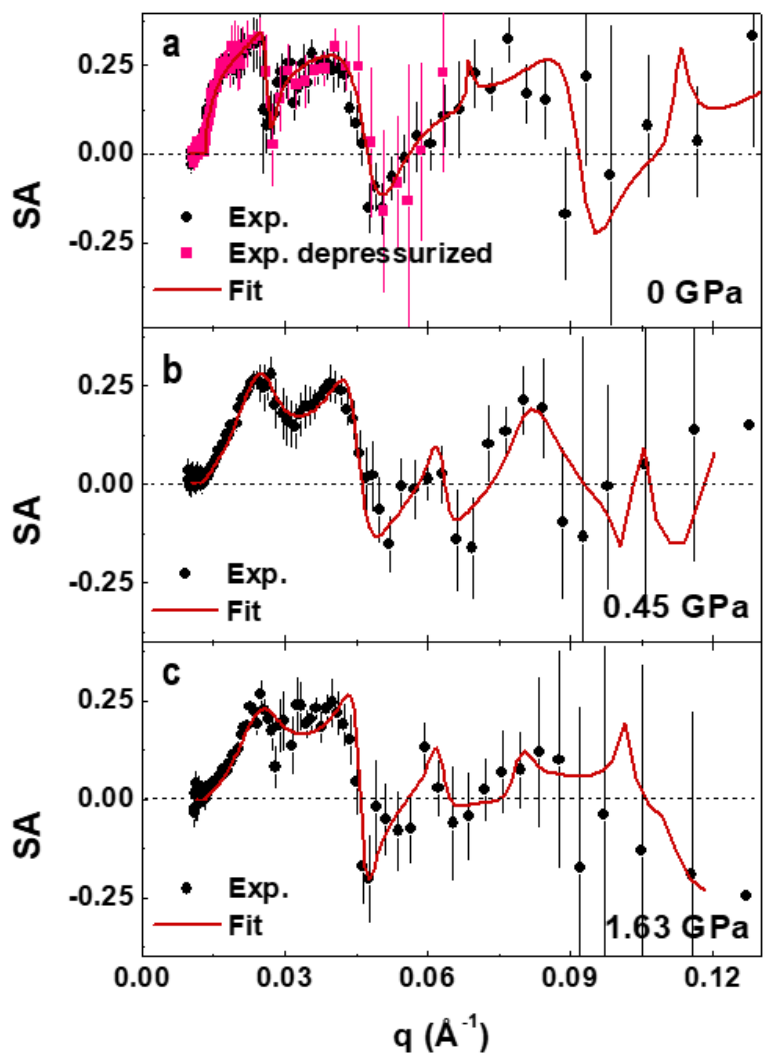

**Figure 4. Suppression of magnetization in the LCO film under hydrostatic pressure.** (**a**) Nuclear (solid lines) and magnetic (shadow areas) SLD depth profiles as a function of the distance from the STO substrate when the hydrostatic pressure increases from 0 to 1.63 GPa. (**b**) Pressure dependent magnetization of the LCO film. The square symbols represent the magnetization of the LCO film bulk, and the circle symbols indicate the magnetization of the LCO interface layer in proximity to the STO substrate/capping layer. The dashed lines are linear fits to the magnetization data. (**c**)-(**e**) Schematic energy-level diagrams of $Co^{3+}$ ion with LS ($t_{2g}^6 e_g^0$, $S = 0$), IS ($t_{2g}^5 e_g^1$, $S = 1$), and HS ($t_{2g}^4 e_g^2$, $S = 2$) configurations, respectively.

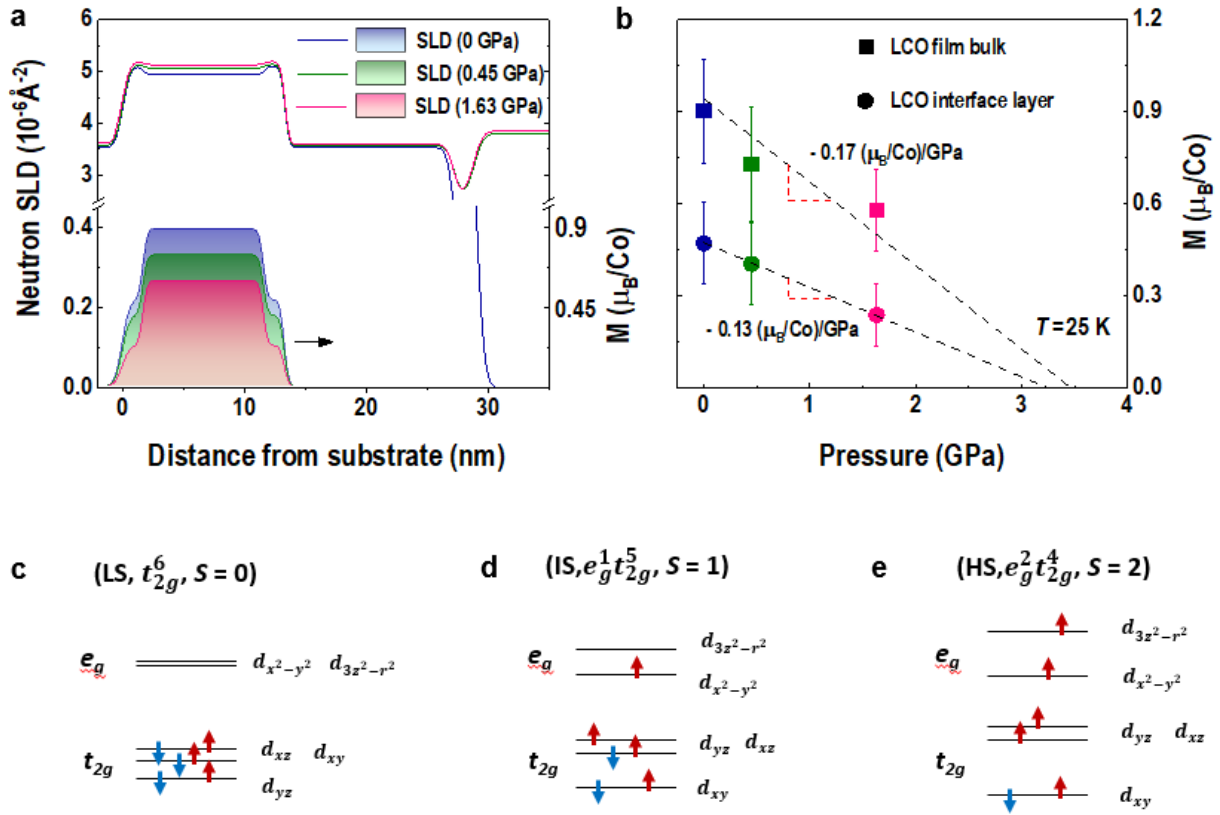